\documentclass[aps, reprint, superscriptaddress,showpacs, floatfix]{revtex4-1}
\usepackage{amsmath}
\usepackage{amssymb}
\usepackage{bm}
\usepackage{graphicx}

\newcommand{\mvec}[1]{\ensuremath{\mathbf{#1}}} 
\newcommand{\mvect}[2]{\ensuremath{\mathbf{#1}_\mathrm{#2}}} 
\newcommand{\Heff}{\ensuremath{\mvec{H}_\mathrm{eff}}} 
\newcommand{\RBar}{\ensuremath{\mvec{B}_\mathrm{Bar}}}
\newcommand{\Hpar}{\ensuremath{\mvec{H}_\mathrm{eff}^\parallel}} 
\newcommand{\Hperp}{\ensuremath{\mvec{H}_\mathrm{eff}^\perp}} 
\newcommand{\Hb}{\ensuremath{\mvec{H}_\mathrm{eff}^\mathrm{B}}} 
\newcommand{\tauperp}{\bm{\tau}_\perp}
\newcommand{\taupar}{\bm{\tau}_\parallel}

\begin{document}

\title{Phenomenological description of the nonlocal magnetization relaxation in magnonics, spintronics, and domain-wall dynamics}

\author{Weiwei Wang}
\affiliation{Engineering and the Environment, University of Southampton, Southampton, UK}
\author{Mykola Dvornik}
\affiliation{DyNaMat Lab, Ghent University, Gent, Belgium}
\affiliation{Physics Department, University of Gothenburg, 412 96, Gothenburg, Sweden}
\author{Marc-Antonio Bisotti}
\affiliation{Engineering and the Environment, University of Southampton, Southampton, UK}
\author{Dmitri Chernyshenko}
\affiliation{Engineering and the Environment, University of Southampton, Southampton, UK}
\author{Marijan Beg}
\affiliation{Engineering and the Environment, University of Southampton, Southampton, UK}
\author{Maximilian Albert}
\affiliation{Engineering and the Environment, University of Southampton, Southampton, UK}
\author{Arne Vansteenkiste}
\affiliation{DyNaMat Lab, Ghent University, Gent, Belgium}
\author{Bartel V. Waeyenberge}
\affiliation{DyNaMat Lab, Ghent University, Gent, Belgium}
\author{Andriy N. Kuchko}
\affiliation{Physical and Technical Department, Donetsk National University, Donetsk, Ukraine}
\affiliation{Institute of Magnetism of NAS of Ukraine, 36b Vernadskogo Avenue, Kiev, 03142, Ukraine}
\author{Volodymyr V. Kruglyak}
\affiliation{School of Physics, University of Exeter, Exeter, UK}
\author{Hans Fangohr}
\affiliation{Engineering and the Environment, University of Southampton, Southampton, UK}

\begin{abstract}
A phenomenological equation called Landau-Lifshitz-Baryakhtar (LLBar) equation, which could be viewed as 
the combination of Landau-Lifshitz (LL) equation and an extra ``exchange damping" term, was derived by 
Baryakhtar using Onsager's relations. We interpret the origin of this ``exchange damping" as nonlocal damping 
by linking it to the spin current pumping. The LLBar equation is investigated numerically and analytically for 
the spin wave decay and domain wall motion. Our results show that the lifetime and propagation length of 
short-wavelength magnons in the presence of nonlocal damping could be much smaller than those given by LL equation. 
Furthermore, we find that both the domain wall mobility and the Walker breakdown field are strongly influenced by the nonlocal damping.
\end{abstract}


\maketitle

\section{Introduction}
The genuine complexity of magnetic and spintronic phenomena occurring in magnetic samples and devices imposes 
both fundamental and technical limits on the applicability of quantum-mechanical and atomistic theories to their modeling.  
To a certain degree, this challenge can be circumvented by exploiting phenomenological theories based on the continuous 
medium approximation.  The theories operate with the magnetization (i.e. the magnetic moment density) and the effective 
magnetic field as generalized coordinates and forces respectively \cite{Landau1935, Akhiezer1968}. 
The effective magnetic field is defined in terms of various magnetic material parameters, which are determined 
by fitting theoretical results to experimental data, and at least in principle, can be calculated using the 
quantum-mechanical or atomistic methods. However, solving the phenomenological models analytically is still 
a formidable task in the majority of practically important cases.  The difficulty is primarily due to the 
presence of the long range magneto-dipole interaction and associated non-uniformity of the ground state 
configurations of both the magnetization and effective magnetic field.  Hence, the phenomenological models 
are solved instead numerically, using either finite-difference or finite-element methods realized in a number of 
micromagnetic solvers \cite{Porter1999, Scholz2003, Fischbacher2007, Berkov2008,Vansteenkiste2014}.

Traditionally, the software for such numerical micromagnetic simulations of magnetization dynamics is based on 
solving the Landau-Lifshitz equation \cite{Landau1935} with a transverse magnetic relaxation term, either in 
the original (Landau) \cite{Landau1935} or "Gilbert" \cite{Gilbert2004}  form. 
Over time, dictated by the experimental and technological needs, the solvers have been modified to include 
finite temperature effects \cite{prb14937} and additional contributions to the magnetic energy (and therefore 
to effective magnetic field) \cite{Shu2004}.  The recent advances in spintronics and magnonics have led to 
the implementation of various spin transfer torque terms \cite{Zhang2004,Thiaville2005} and 
periodic boundary conditions \cite{Lebecki2008, Wang2010, Krueger2013}.  Furthermore, the progress in experimental 
investigations of ultrafast magnetization dynamics \cite{Kirilyuk2010} has exposed the need to account for the variation 
of the length of the magnetization vector in response to excitation by femtosecond optical pulses, leading to inclusion of 
the longitudinal relaxation of the magnetization within the formalism of numerical micromagnetics \cite{Au2013}.  
Provided that a good agreement between the simulated and measured results is achieved, a microscopic (i.e. quantum-mechanical or atomistic) 
interpretation of the experiments can then be developed.

The described strategy relies on the functional completeness of the phenomenological model.  
For instance, a forceful use of incomplete equations to describe phenomena originating from terms 
missing from the model may result in false predictions and erroneous values of fitted parameters, 
and eventually in incorrect conclusions.  The nature of the magnetic relaxation term and associated 
damping constants in the Landau-Lifshitz equation is of paramount importance both fundamentally and technically.  
It is this term that is responsible for establishment of equilibrium both within the magnetic sub-system and 
with its environment (e.g. electron and phonon sub-systems), following perturbation by magnetic fields, spin currents, 
and/or optical pulses \cite{Kirilyuk2010}.  Moreover, it is the same term that will eventually determine 
the energy efficiency of any emerging nano-magnetic devices, including both those for data storage \cite{Terris2005} 
and manipulation \cite{Kruglyak2010}.

In this report, we demonstrate how the phenomenological magnetic relaxation term derived by Baryakhtar 
to explain the discrepancy between magnetic damping constants obtained from ferromagnetic resonance (FMR) 
and magnetic domain wall velocity measurements in dielectrics \cite{Baryakhtar1998, Baryakhtar1986, Baryakhtar2013}  
can be applied to magnetic metallic samples. We show that the Landau-Lifshitz equation 
with Baryakhtar relaxation term (Landau-Lifshitz-Baryakhtar or simply LLBar equation) contains 
the Landau-Lifshitz-Gilbert (LLG) equation as a special case, while also naturally including the contribution 
from the nonlocal damping in the tensor form of Zhang and Zhang \cite{Zhang2009} and De Angeli \cite{DeAngeli2009}.  
The effects of the longitudinal relaxation and the anisotropic transverse relaxation on the magnetization dynamics 
excited by optical and magnetic field pulses, respectively, in continuous films and magnetic elements 
were discussed e.g. in Ref. [\citenum{Au2013, Dvornik2013, Yastremsky2014}].  
So, here we focus primarily on the manifestations of the Baryakhtar relaxation in problems 
specific for magnonics \cite{Kruglyak2010} and domain wall dynamics \cite{BurkardHillebrands2006, Weindler2014}.  
This is achieved by incorporating the LLBar equation within the code of the Object Oriented Micromagnetic 
Framework (OOMMF) \cite{Porter1999},  probably the most popular micromagnetic solver currently available, 
and by comparing the results of simulations with those from simple analytical models.  
Specifically, we demonstrate that the Baryakhtar relaxation leads to increased damping of short wavelength 
spin waves and to modification of the domain wall mobility, the latter being also affected by the longitudinal relaxation strength.

The paper is organized as follows. In Sec. II, we review and interpret the Baryakhtar relaxation term.
In Sec. III, we calculate and analyze the spin wave decay in a thin magnetic nanowire. 
In Sec. IV, we simulate the the suppression of standing spin waves in thin film. 
In Sec. V, we analyze the domain wall motion driven by the external field and compare the relative strength 
of contributions from the longitudinal and nonlocal damping.  We conclude the discussion in Sec. VI.  
 
\section{Basic equations}
In the most general case, the LLBar equation can be written as \cite{Baryakhtar1998, Dvornik2013}
\begin{equation}\label{eq_llbar1}
\frac{\partial \mvec{M}}{\partial t}=
-\gamma \mvec{M} \times \Heff + \mvec{R}
\end{equation}
where $\gamma (>0)$ is the gyromagnetic ratio and the relaxation term $\mvec{R}$ is
\begin{equation}
\mvec{R} =\hat{\Lambda}_r \cdot \Heff - \hat{\Lambda}_{e,sp}  \frac{\partial^2{\Heff}}{\partial x_s \partial x_p}.
\end{equation}
Here and in the rest of the paper, the summation is automatically assumed for repeated indices.  
The two relaxation tensors $ \hat{\Lambda}_r$  and $ \hat{\Lambda}_e$  describe relativistic and exchange contributions, 
respectively, as originally introduced in Ref.~[\citenum{Baryakhtar1986}].

To facilitate comparison with the LLB equation as written in Ref.~[\citenum{Atxitia2011}], the magnetic 
interaction energy of the sample is defined as  
\begin{equation}\label{eq_exch}
w = w_{\mu} + \frac{\mu_0}{8\chi} \frac{(M^2-M_e^2)^2}{ M_e^2},
\end{equation}
where $M_e$ is the equilibrium magnitude of the magnetization vector at a given temperature and 
zero micromagnetic effective field, i.e. the effective field derived from the micromagnetic energy density 
$w_\mu$, as used in standard simulations at constant temperature under condition  
$|\mvec{M}|=M_e=\mathrm{const}$ (i.e. with only the transverse relaxation included).  
The second term in right-hand side of Eq.~(\ref{eq_exch}) describes the energy density induced by the small deviations of the magnetization length 
from its equilibrium value $M_e$ at the given temperature, i.e., $|M^2-M_e^2|\ll M_e^2$, and $\chi$ is the longitudinal magnetic susceptibility. 
Therefore, the associated effective 
magnetic field is
\begin{equation}\label{eq_heff0}
\Heff=-\frac{1}{\mu_0} \frac{\delta w}{\delta \mvec{M}} 
= \mvec{H}_{\mu} + \frac{1}{2\chi} (1-n^2) \mvec{M} 
\end{equation}
where $\mvec{n}=\mvec{M}/M_e$, $\mvec{H}_\mu$  is the effective magnetic field associated to $w_\mu$. 
Hereafter we assume that our system is in contact with the heat bath, so that the equilibrium temperature and
associated value of $M_e$ and $\chi$ remain constant irrespective of the magnetization dynamics.

In accordance with the standard practice of both micromagnetic simulations and analytical calculations, to solve LLBar equations (\ref{eq_llbar1}-\ref{eq_heff0}), 
one first needs to the corresponding static equations obtained by setting the time derivatives to zero and 
thereby to derive the spatial distribution of the magnetization in terms of both its length and direction.  
We note that, in general (e.g. as in the case of a domain wall), the resulting distribution of the longitudinal effective field 
and therefore also of the equilibrium magnetization length is nonuniform, so that the length is not generally equal to $M_e$.  
With the static solution at hands, the dynamical problem is solved so as to find the temporal evolution of the magnetization length 
and direction following some sort of a perturbation.  Crudely speaking, the effect of the relaxation terms is that, at each moment of time, 
the magnetization direction relaxes towards the instantaneous direction of the effective magnetic field, 
while the magnetization length relaxes towards the value prescribed by the instantaneous longitudinal effective magnetic field.  
The effective field itself varies with time, which makes the problem rather complex.  However, this is the same kind of complexity as the one that 
has always been inherent to micromagnetics.  The account of the longitudinal susceptibility within the LLBar equation only brings 
another degree of freedom (the length of the magnetization) into the discussion.  
One should note however that the longitudinal susceptibility has a rather small value at low temperature and 
so its account is only required at temperatures of the order of the Curie temperature. 

We neglect throughout the paper any effects due to the anisotropy of relaxation, which could be associated e.g. with the 
crystalline structure of the magnetic material \cite{Baryakhtar1998,Dvornik2013}. This approximation is justified for 
polycrystalline and amorphous soft magnetic metals, as has been confirmed by simulations presented in Ref. [\citenum{Dvornik2013}]. 
Hence, we represent the relaxation tensors as
 $\hat{\Lambda}_r = \lambda_r \hat{I}$
and $\hat{\Lambda}_e = \lambda_e \hat{I}$ where parameters $\lambda_r$ and $\lambda_e$ are the
relativistic and exchange
relaxation damping constants and $\hat{I}$ is the unit tensor. Then, Eq.~(\ref{eq_llbar1}) is reduced to
\begin{equation}\label{eq_B2}
\partial_t \mvec{M}=
-\gamma \mvec{M} \times \mvec{H}_{\mathrm{eff}}
+\lambda_r  \mvec{H}_{\mathrm{eff}} - \lambda_e \nabla^2 \mvec{H}_{\mathrm{eff}}.
\end{equation}

We separate the equations describing the dynamics and relaxation of the length and direction of the magnetization vector.  
Representing the latter as a product of its magnitude and directional unit vector  $\mvec{M} = M \mvec{m}$, we can write 
\begin{equation}\label{eq_split}
M \frac{\partial \mvec{m}}{\partial t} + \mvec{m} \frac{\partial M}{\partial t} =
-\gamma \mvec{M} \times \Heff + \mvec{R}.
\end{equation}
We multiply this equation by $\mvec{m}$ to obtain,
\begin{equation}\label{eq_M}
\frac{\partial M}{\partial t} = \mvec{m} \cdot \mvec{R}.
\end{equation} 
Then, subtracting the product of equation (\ref{eq_M}) and $\mvec{m}$ from equation (\ref{eq_split}), we obtain 
\begin{equation}\label{eq_LLR}
\frac{\partial \mvec{m}}{\partial t}  =
-\gamma \mvec{m} \times \Heff + \frac{1}{M} \mvec{R}_\perp
\end{equation}
where $\mvec{R}_\perp=-\mvec{m}\times (\mvec{m} \times \mvec{R})$. In the rest of the paper, 
we will use $\mvec{A}_\perp \equiv (\mvec{A})_\perp \equiv \mvec{A} - (\mvec{A} \cdot \mvec{m}) \mvec{m}$ 
to represent the component of the vector $\mvec{A}$ that is perpendicular (transverse) to vector $\mvec{m}$. 
Note that only the perpendicular component of the torque contributes to $\partial_t \mvec{m} \equiv \partial{\mvec{m}}/\partial t$.
For given temperature, $M_e$ is constant and we can define $\alpha = \lambda_r/(\gamma M_e)$.
In the limiting case of $\chi \rightarrow 0$, $M \rightarrow M_e$ and thus $\alpha$ is recognized as
the Gilbert damping constant from the LLG equation.
Let us now consider the case of  $ \hat{\Lambda}_e \neq 0$.  The corresponding contribution to the relaxation term, 
which we denote here as $\RBar$, can be written as
\begin{equation}
\RBar =  - \lambda_e \nabla^2 \mvec{H}_{\mathrm{eff}} \equiv - \partial_i  \mvec{j}_i,
\end{equation}
where $\partial_i \equiv \partial/\partial x_i$ and the quantity $\mvec{j}_i = -\lambda_e  \partial_i \Heff$ has 
the form of some magnetization current density (magnetization flux). 

For the following, it is useful to split the effective field into its perpendicular (relative to \mvec{m}) part 
($\Hperp$, ``perpendicular field") and  parallel part ($\Hpar$, ``parallel field"), i.e., $\Heff=\Hperp+\Hpar$,  
and then to consider the associated magnetic fluxes and torques separately.  
The magnetic flux of $\mvec{j}_{\parallel,i}=-\lambda_e \partial_i \Hpar$ and then the contribution of  
the associated torque $\taupar=-\partial_i \mvec{j}_{\parallel,i} $ onto  $\mvec{m}$ is 
\begin{equation}
(\taupar)_\perp = -2 \lambda_e \partial_i \Hpar \partial_i \mvec{m} - \lambda_e \Hpar (\nabla^2 \mvec{m})_\perp.
\end{equation}
The perpendicular field can be represented as 
\begin{equation}\label{eq_Heff}
\Hperp = \frac{1}{\gamma M^2} \left [ \mvec{M} \times \frac{\partial \mvec{M}}{\partial t} \right] + O (\mvec{R}) \approx  \frac{1}{\gamma } \left [ \mvec{m} \times 
\partial_t \mvec{m}\right]. 
\end{equation}
So, we can write for the magnetization flux associated with the perpendicular field
\begin{equation}\label{eq_jperp}
\mvec{j}_{\perp,i} = - (\lambda_e/\gamma) \partial_i (\mvec{m} \times \partial_t \mvec{m}).
\end{equation}
The right-hand side of Eq.~(\ref{eq_jperp}) could be regarded as the torque generated by spin current pumping since 
$\mvec{m} \times \partial_t \mvec{m}$ can be considered as the exchange spin current \cite{Tserkovnyak2009},   
and then for the associated perpendicular torque $\tauperp$, we obtain, 
 \begin{equation}
 \tauperp = -\partial_i \mvec{j}_{\perp,i}  = - \sigma M_e \partial_i \partial_i (\mvec{m} \times \partial_t \mvec{m} ),
 \end{equation}
 where we have introduced variable $\sigma=\lambda_e/(\gamma M_e)$.
 We show that the torque $(\bm{\tau}_\perp)_\perp$ could be written as  (see Appendix \ref{app_llb} for details)
 \begin{equation}
 (\bm{\tau}_\perp)_\perp = M_e \left[
 \mvec{m}\times (\mathcal{D} \cdot \partial_t \mvec{m}) 
 - \sigma \mvec{m} \times \nabla^2  \partial_t \mvec{m} \right ]
\end{equation} 
where $\mathcal{D}$ is a $3\times 3$ tensor \cite{Zhang2009, Fahnle2013},
\begin{equation}\label{eq_Dtensor}
\mathcal{D}_{\alpha\beta} = 2 \sigma (\mvec{m}\times \partial_i \mvec{m})_\alpha (\mvec{m}\times \partial_i \mvec{m})_\beta
- \sigma (\partial_i \mvec{m} \cdot \partial_i \mvec{m}) \delta_{\alpha \beta} .
\end{equation}
In the limit of $\chi\rightarrow 0$, we assume $\Hpar=0$ and therefore obtain
\begin{eqnarray}\label{eq_llb}
\partial_t \mvec{m} = -\gamma \mvec{m} \times \mvect{H}{eff} 
-\gamma \alpha \mvec{m} \times (\mvec{m} \times \mvect{H}{eff}) \nonumber\\
 + \mvec{m}\times (\mathcal{D} \cdot \partial_t \mvec{m}) 
 - \sigma \mvec{m} \times \nabla^2  \partial_t \mvec{m}.
\end{eqnarray} 
At the same time, Eq.~(\ref{eq_LLR}) can then be written as 
\begin{equation} \label{eq_bterm}
\frac{\partial \mvec{m}}{\partial t}= -\gamma \mvec{m} \times \mvec{H}_{\mathrm{eff}} 
- \gamma \mvec{m} \times ( \mvec{m} \times \Hb),
\end{equation}
where 
\begin{equation}\label{eq_Hb}
\Hb = \alpha \mvect{H}{eff} - \sigma \nabla^2 \mvec{H}_\mathrm{eff}^{\perp},
\end{equation}
and $\mvec{H}_\mathrm{eff}^{\perp}$ is the transverse component of the effective field. 
The first term in Eq.~(\ref{eq_Hb}) is kept as $\mvec{H}_\mathrm{eff}$ since 
$\mvec{m} \times \mvec{H}_\mathrm{eff} = \mvec{m} \times  \mvec{H}_\mathrm{eff}^{\perp}$.
In practice, we use Eq.~(\ref{eq_bterm}) rather than Eq.~(\ref{eq_llb}) for numerical implementation.
As shown in Eq.~(\ref{eq_llb}) the damping terms contain both the 
form $-\mvec{m} \times \nabla^2  \partial_t \mvec{m}$ \cite{Hankiewicz2008, Tserkovnyak2009}
and tensor form  $\mvec{m}\times (\mathcal{D} \cdot \partial_t \mvec{m})$~\cite{Zhang2009}. 
Hence, we conclude that the exchange damping can be explained as the nonlocal damping, and Eq.~(\ref{eq_bterm}) 
is the phenomenological equation to describe the nonlocal damping.

The intrinsic Gilbert damping is generally considered to have the relativistic origin~\cite{Landau1935,Hickey2009}.  
Phenomenologically, the Gilbert damping is local and the damping due to the nonuniform magnetization 
dynamics being ignored~\cite{Gilbert2004}. The exchange relaxation term in the LLBar equation 
describes the nonlocal damping due to the nonuniform effective field.
Despite the complexity of various damping mechanisms, the spin current $\mvec{j}$ 
in conducting ferromagnets can be calculated, e.g. using the time-dependent Pauli equation within the s-d model. 
The spin current is then given by $\mvec{j}_i=(g\mu_B \hbar G_0/4e^2)(\partial_t \mvec{m} \times \partial_i \mvec{m})$,
where $G_0$ is the conductivity~\cite{Zhang2009}, and thus the nonlocal damping of the tensor form can be obtained~\cite{Zhang2009,Fahnle2013}. 
As we can see from Appendix \ref{app_llb}, this spin current densities $\mvec{j}_i$ and $\mvec{j}_i^a$ have the same form, and therefore, 
we can establish that $\sigma \sim g\mu_B \hbar G_0/4e^2 M_e$. The spin current component $\mvec{j}_i^b$ (see Appendix \ref{app_llb}) gives the term 
$-\mvec{m} \times \nabla^2  \partial_t \mvec{m}$~\cite{Tserkovnyak2009}, and  the value of $\sigma$ can be therefore interpreted as
$\sigma \sim (\gamma/\mu_0 M_e) (\hbar/2)^2 n_e \tau_\mathrm{sc}/m^*$, where $n_e$ is the conduction electron density, $m^*$ the effective
mass and $\tau_\mathrm{sc}$ is the transverse spin scattering time~\cite{Nembach2013}.

It is of interest to compare Eq.~(\ref{eq_B2}) with Landau-Lifshitz-Bloch (LLB) equation \cite{Atxitia2011}, 
which could be written as 
\begin{equation}\label{eq_LLB}
\frac{\partial \mvec{n}}{\partial t}=
- \gamma  \mvec{n} \times \mvec{H}_{\mathrm{eff}}
+  \frac{\gamma \alpha_\parallel}{n^2} [\mvec{n} \cdot \mvect{H}{eff}]  \mvec{n}
- \frac{\gamma \alpha_\perp}{n^2} \mvec{n} \times ( \mvec{n} \times \mvect{H}{eff})
\end{equation}  
where $\mvec{n}=\mvec{M}/M_e(T)$ is the reduced magnetization and $M_e(T)$ is the equilibrium magnetization value at temperature $T$. 
The effective field $\mvect{H}{eff}$ contains the usual micromagnetic contributions $\mvect{H}{int}$ as well as the contribution from 
the temperature, 
\begin{equation}\label{eq_H}
\mvect{H}{eff} = \mvect{H}{int} + \frac{m_e}{2 \tilde{\chi}_\parallel}(1-n^2) \mvec{n}
\end{equation}
where $m_e=M_e(T)/M_e(0)$ and $\tilde{\chi}_\parallel=\partial m/\partial H$ with $m=M/M_e(0)$ \cite{Atxitia2011}.  
By substituting  Eq.~(\ref{eq_H}) into Eq.~(\ref{eq_LLB}), one arrives at
\begin{eqnarray}
\frac{\partial \mvec{n}}{\partial t}=
- \gamma  \mvec{n} \times \mvect{H}{int}
+  \gamma \alpha_\parallel   (\mvect{H}{int})_\parallel
+ \gamma \alpha_\perp   (\mvect{H}{int})_\perp  \nonumber \\
+ \frac{\alpha_\parallel \gamma m_e}{2 \tilde{\chi}_\parallel }(1-n^2) \mvec{n}.
\end{eqnarray}  
Meanwhile, if we neglect the $\lambda_e$ term in Eq.~(\ref{eq_B2}) and insert the effective field Eq.~(\ref{eq_exch}) into Eq.~(\ref{eq_B2}), 
we obtain
\begin{equation}\label{eq_llbar3}
\frac{\partial \mvec{n}}{\partial t}=
- \gamma  \mvec{n} \times \mvect{H}{int}
+  \gamma \lambda_r  \mvect{H}{int}
+ \frac{\lambda_r \gamma }{2\chi}(1-n^2) \mvec{n}.
\end{equation}  
As we can see, Eq.~(\ref{eq_llbar3}) is a special case of LLB equation with the assumption 
that $\alpha_\perp=\alpha_\parallel=\lambda_r/(\gamma M_e)$ and $\chi= M_e(0)\tilde{\chi}_\parallel$. 
However, the LLB equation does not contain the $\lambda_e$-term (nonlocal damping term) which is the main focus in this work.

\section{Spin wave decay} \label{section_sw}
To perform the micromagnetic simulation for the spin wave decay, we have implemented 
Eq.~(\ref{eq_bterm}) as an extension for the finite difference micromagnetics 
package OOMMF. 
A new variable $\beta$ for the exchange damping is introduced with 
$\sigma=\beta G$, where $G$ is a coefficient to scale $\beta$ to 
the same order as $\alpha$.  In practice, $G$ was chosen to be $G=A/(\mu_0 M_e^2)$.

The simulation geometry  has dimensions  $L_x=2002\,\textrm{nm}$,  $L_y=2\,\textrm{nm}$ 
and  $L_z=2\,\textrm{nm}$, and the cell size is $1\times 2\times 2\, \mathrm{nm}^3$.  
The magnetization aligns along the $\mvec{e}_x$ direction for the equilibrium state and the  parameters 
are typical of Permalloy: the exchange constant $A = 1.3 \times 10^{-11}\,\mathrm{J/m}$, 
the saturation magnetization $M_e = 8.6\times 10^5\,\mathrm{A/m}$ and the Gilbert damping damping coefficient  $\alpha=0.01$.
The spin waves are excited locally in the region $0\leq x \leq 2\,\textrm{nm}$, and to prevent 
the spin wave reflection the damping coefficient is increased linearly \cite{Han2009a} from 0.01 at 
$x=1802\,\textrm{nm}$ to 0.5 at $x=2002\,\textrm{nm}$. 

\begin{figure}[htbp]
\includegraphics[width=0.42\textwidth]{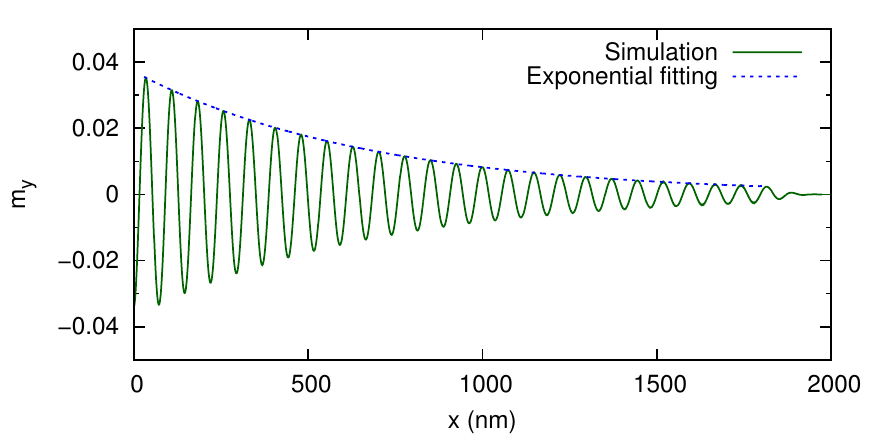} 
\caption{The spin wave amplitude decay along the rod, for a spin wave was excited locally 
by applying a microwave $\mvec{H}=H_{0}\sin(2 \pi f t) \mvec{e}_y$ of frequency $f=30\,\mathrm{GHz}$ 
and amplitude $H_0=1000\,\mathrm{Oe}$ in the region $0\leq x \leq 2\,\mathrm{nm}$.  
The data were fitted using Eq.~(\ref{eq_sw_d}) with $\beta=0.02$ and $\alpha=0.01$.}
\label{fig_sw_decay}
\end{figure}
Figure~\ref{fig_sw_decay} illustrates the spin wave amplitude decay along the rod.  
The $y$ component of magnetization unit vector $m_y$ data for $30 \leq x \leq 1800 $ nm 
were fitted using (\ref{eq_sw_d}) to extract the wave vector $k$ and the decay constant $\lambda$, 
and good agreement is observed due to the effective absence of spin wave reflection.
We use data after having computed the time development of the magnetization for 4 ns to reach a steady state.
 The injected spin wave energy is absorbed efficiently enough within the 
right 200 nm of the rod due to the increased damping.

To analyze the simulation data, we exploit the uniform plane wave assumption with its exponential
amplitude decay due to energy dissipation,
 i.e. magnetization with the form $e^{i(\mathrm{k}x -\omega t)} e^{-\lambda x}$, where $\lambda$ is the 
 characteristic parameter of the spin wave damping.
For a small amplitude spin wave propagation we have \cite{Seo2009}
\begin{equation}\label{eq_sw_d}
\mvec{m}= \mvec{e}_\mathrm{x} + \mvec{m}_0 e^{i(k x - \omega t)} e^{-\lambda x}
\end{equation}
where $|\mvec{m}_0| \ll 1$,
and the effective field of the long rod can be expressed as 
\begin{equation}\label{eq_heff}
\mvec{H}_\mathrm{eff}=H_s m_x \mvec{e}_x + D \nabla^2 \mvec{m},
\end{equation}
where the `easy axis' anisotropy field $H_s  m_x \mvec{e}_x$ originates from the demagnetizing field, 
and the constant $D$ measures the strength of the exchange field,
\begin{equation}
H_s= \frac{2K}{\mu_0 M_e}=\frac{1}{2} M_e, \qquad  D= \frac{2 A}{\mu_0 M_e}.
\end{equation}
To test the spin wave decay for this system, a sinusoidal field $\mvec{H}=H_{0}\sin(2 \pi f t) \mvec{e}_y$ 
was applied to the rod in the region $0\leq x \leq 2\,\mathrm{nm}$ to generate spin waves.

\begin{figure}[htbp]
\includegraphics[width=0.4\textwidth]{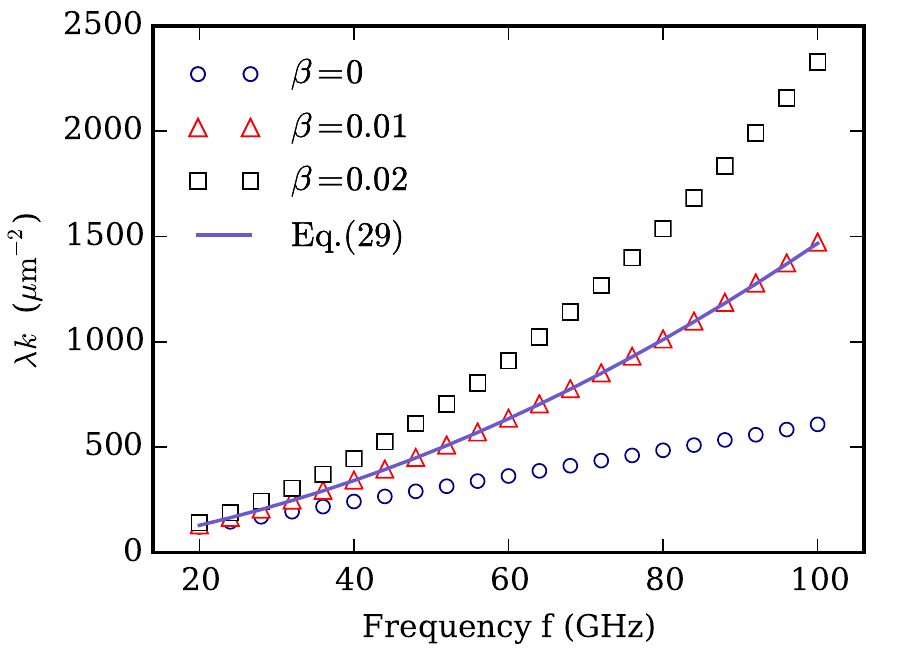} 
\caption{
The spin wave decay constant--wave vector product $\lambda k$ as a function of the frequency 
for different $\beta$ values. The slateblue line was drawn using Eq.~(\ref{eq_lambda_k}) for the case $\beta=0.01$.  }
\label{fig_sw_damping}
\end{figure}

Figure~\ref{fig_sw_damping} shows the product of spin wave decay constant $\lambda$ and 
wave vector $k$ as a function of the frequency.  The dependence is linear for the $\beta=0$ case, 
which is agreement with the zero adiabatic spin torque case \cite{Seo2009}.
 The addition of a nonzero $\beta$ term leads to a nonlinear relation, 
 and the amplitude of the spin wave decay constant that is significantly larger than that given by the linear dependence. 
We also performed the simulation for $\chi>0$ case by using Eq.~(\ref{eq_B2})  which shows
that $\beta$ term is the leading factor for this nonlinearity (the relative error is less than 1\% for $\chi=1\times 10^{-3}$).  
To analyze the nonlinear dependence,  we introduce the complex wave vector $\widetilde{k}$,
\begin{equation}\label{eq.complex_k}
\widetilde{k}=k+\lambda i.
\end{equation}  
By linearizing  Eq.~(\ref{eq_bterm})  
and setting the determinant of the matrix to zero we obtained (see Appendix \ref{app_sw} for details):
\begin{equation}\label{eq_sw}
(\omega + \tilde{\omega}_0 + i \tilde{\omega}_1 )(\omega - \tilde{\omega}_0 + i \tilde{\omega}_1)=0,
\end{equation}
where
$\tilde{\omega}_0=\gamma(H_s+D \tilde{k}^2)$ and
$\tilde{\omega}_1=   \alpha \tilde{\omega}_0 +  \beta G \tilde{k}^2 \tilde{\omega}_0$.
The second term of the Eq.~(\ref{eq_sw}) is expected to be equal to zero, i.e., $\tilde{\omega}_1- i \omega + i \tilde{\omega}_0  = 0$.
There are two scenarios to consider: First is the $\beta=0$ case, $k \lambda$ could be extracted by taking 
the imaginary part of $\tilde{k}^2$ at Eq.~(\ref{eq.complex_k}):
\begin{equation}
k \lambda =\frac{1}{2}\,  \mathrm{Im} \left\{   \tilde{k}^2   \right\} = \frac{\alpha \omega}{ 2(1+\alpha^2)\gamma D}.
\end{equation}
The linear dependence of $k\lambda$ as a function of frequency matches the data plotted in Fig. \ref{fig_sw_damping}.
For the $\beta>0$ case, solving Eq. (\ref{eq_sw}) yields in the linear with respect to the damping constants approximation,
\begin{equation}\label{eq_lambda_k}
k \lambda \approx \frac{\omega }{2\gamma D}(\alpha+\beta G k^2)
\end{equation}
where the dispersion relation for the rod is $\omega = \gamma (H_s + D k^2)$.
Eq.~(\ref{eq_lambda_k}) shows there is an extra $k^2$ term associated with the exchange damping term besides 
the linear dependence between $k \lambda$ and $\omega$.
 The slateblue line 
in Fig.~\ref{fig_sw_damping} is plotted using Eq.~(\ref{eq_lambda_k}) 
with $\beta=0.01$ and $\alpha=0.01$, which shows a good approximation for the simulation data.
Besides, this exchange damping could be important in determining the nonadiabatic spin torque. 
We could establish the value of $\beta$ using the existing experimental data, such as the transverse 
spin current data \cite{Nembach2013} gives $\beta \sim 0.1$ which hints the lifetime and propagation 
length of short-wavelength magnons could be much shorter than those given by the LLG equation \cite{Baryakhtar2014}. 

\section{Suppression of standing spin waves}
In the presence of nonlocal damping, the high frequency standing spin waves in the thin films are suppressed \cite{Baryakhtar2014}. 
If the magnetization at the surfaces are pinned, the spin wave resonance can be excited by a uniform alternating magnetic field \cite{Kittel1959}.
With given out-of-plane external field $H_z$ in the $z$-direction, the frequencies of the 
excited spin waves of the film are given by \cite{sw2009},
\begin{equation}
\omega_n = \omega_0 + \omega_M \lambda_\mathrm{ex}^2 \left( \frac{n \pi}{d} \right)^2
\end{equation}
where $d$ is the film thickness, $\omega_0 = \gamma (H_z-M_e)$, $\omega_M=\gamma M_e$ and
$\lambda_\mathrm{ex}=\sqrt{2A/(\mu_0 M_e^2)}$. The excited spin wave modes are labeled by
the integer $n$,  and the odd $n$ has a nonvanishing interaction with the given uniform alternating magnetic field~\cite{Kittel1959}. 

\begin{figure}[htbp]
\includegraphics[width=0.4\textwidth]{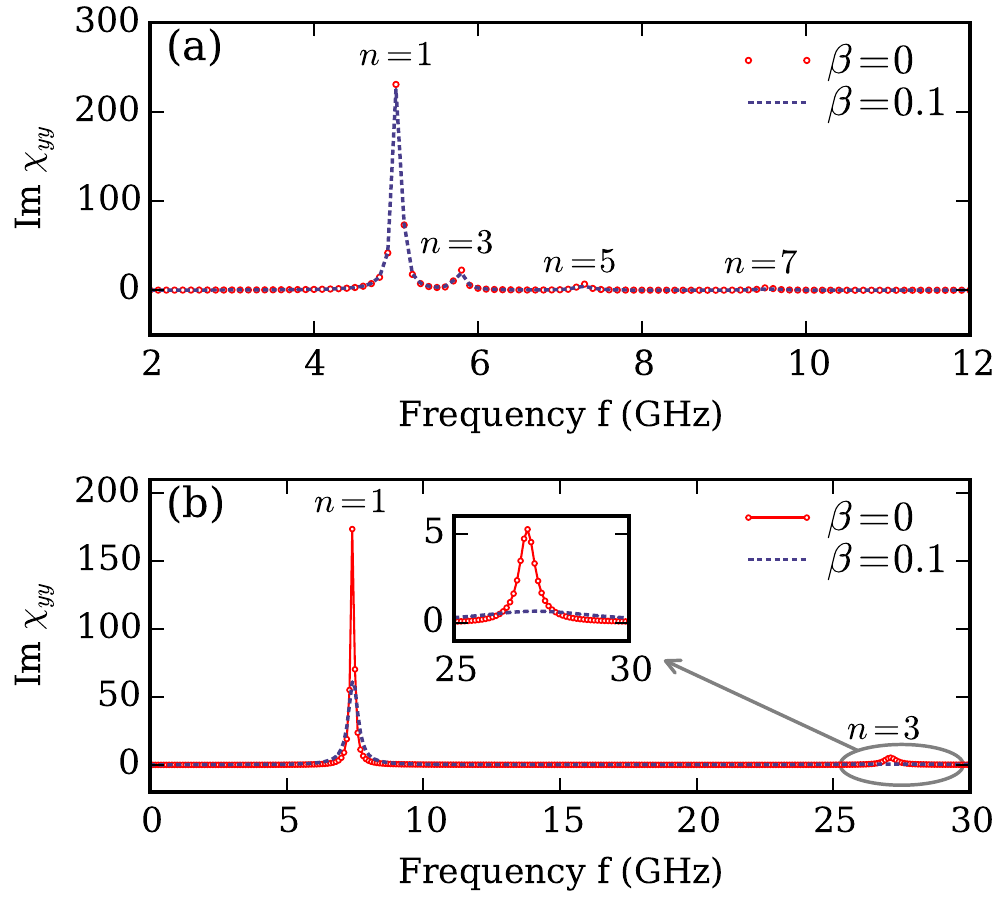} 
\caption{Imaginary parts of the dynamical susceptibility $\chi_{yy}$ of the film for (a) thickness $d=300$ nm, and (b) thickness $d=60$ nm.}
\label{fig_spec}
\end{figure}

To reduce the simulation time, we consider a system with cross-sectional area $4\times4$ nm$^2$ in $xy$-plane and
apply the two-dimensional periodic boundary conditions \cite{Wang2010} to the system. 
We use the Permalloy as the simulation material with external field $H_z=1\times 10^6$ A/m and the cell size is $4\times4\times2$ nm$^3$.
Instead of applying microwaves to the system, we calculate the magnetic absorption spectrum of the film
by applying a sinc-function field pulse $h = h_0 \mathrm{sinc}(\omega_0 t)$ to the system \cite{Dvornik2013a}. 
With the collected average magnetization data, the dynamic susceptibility $\chi$ is 
computed using Fourier transformation~\cite{Liu2008}. For example, the component $\chi_{yy}$ is computed using $m_y$ when
the pulse is parallel to the $y$-axis.

Figure~\ref{fig_spec}(a) shows the imaginary part of the dynamic susceptibility $\chi_{yy}$ for a film with $d=300$ nm.
As we can see, the spin wave of modes $n=1,3,5,\dots$ are excited, and the influence of the ``exchange damping" is small.
However, the presence of the ``exchange damping" suppresses the spin wave excitation ($n>1$ mode) significantly 
for the film with thickness $d=60$ nm, as shown in Fig.~\ref{fig_spec}(b). The reason is because the damping of the 
standing spin waves is proportional to $k^4$ in the presence of exchange damping~\cite{Baryakhtar2014}.

\section{Domain wall motion}\label{section_dw}
\begin{figure}[tbhp]
\begin{center}
\includegraphics[width=0.4\textwidth]{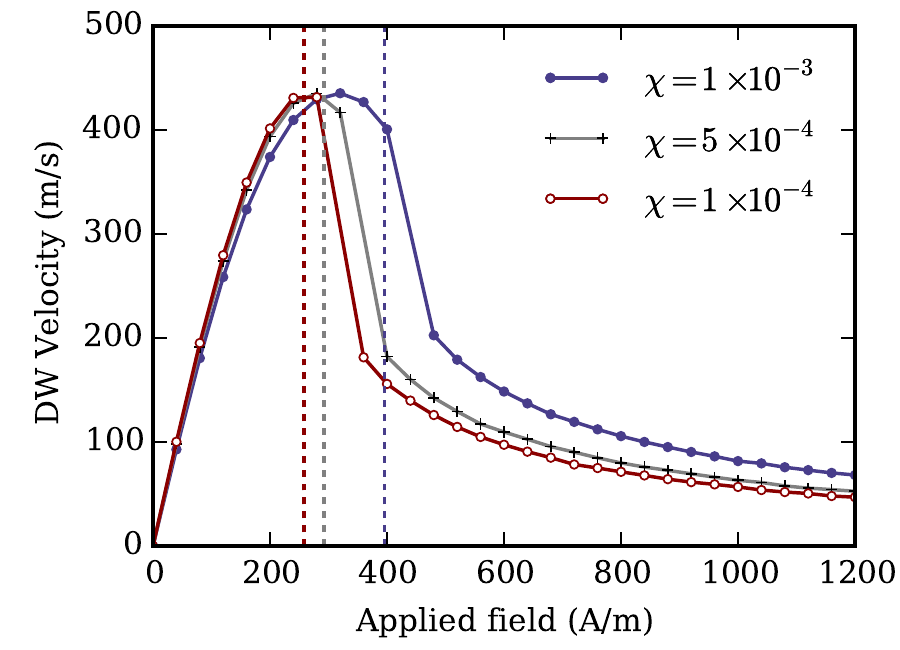} 
\caption{Simulations results of domain wall velocities for various susceptibilities. The parameters used are: 
$\alpha=0.001$, $\beta=0$, $N_y=0.2$ and $N_z=0.8$. The vertical dash lines are the breakdown fields computed 
using Eq.~(\ref{eq.Hk}).}
\label{fig_dw}
\end{center}
\end{figure}

We implemented Eq.~(\ref{eq_B2}) in a finite element based micromagnetic framework
to study the effect of parallel relaxation process on domain wall motion. 
The simulated system for the domain wall motion is a one-dimensional (1D) mesh with length 20000 nm and discretization size  
$4\, \mathrm{nm}$, a head-to-head domain wall is initialized with its center near $x=500$ nm.
In this section, the demagnetizing fields are simplified as $\mvect{H}{d}=-N \mvec{M}$ and the demagnetizing factors are 
chosen to be $N_x=0$, $N_y=0.2$ and $N_z=0.8$, respectively.  The domain wall
moves under the applied field for $50$ ns and the domain wall velocities at different external field strengths are computed. 
Figure~\ref{fig_dw} shows the simulation results of domain wall motion under external fields for different susceptibilities 
without consideration of exchange damping, i.e., $\beta=0$. 
For Nickel and Permalloy, the longitudinal susceptibility is around $10^{-4}$ at room temperature and increases with the temperature up until the Curie point~\cite{Atxitia2011}.
We find that the longitudinal susceptibilities have no influence on the maximum velocity but change the Walker breakdown field $H_w$ significantly.
The domain wall velocity in the limit $\chi \rightarrow 0$ is almost the same with the case of $\chi=10^{-4}$, which
could be explained by the relation that the difference proportional to the ratio of $(\chi/\alpha)^2$ in Eq.~(\ref{eq.Hk}).

To investigate the effect of longitudinal magnetic susceptibility $\chi$ and  exchange relaxation  damping  $\sigma$ 
on the domain wall motion, we use the remainder of this section for analytical studies.
We start from the constant saturation magnetization of one-dimensional domain wall model, 
such as the 1D head-to-head wall~\cite{Nakatani2005}. 
The static 1D domain wall profile can be expressed as 
\begin{equation}
m_x=-\tanh \left(\frac{x-q}{\Delta}\right), \quad
m_t=\mathrm{sech} \left(\frac{x-q}{\Delta}\right)
\end{equation}
where $m_t$ is the perpendicular component of the unit magnetization vector, $\Delta$ is 
the wall width parameter and $q$ is the position of the domain wall center.

We consider the case that the system is characterized by two anisotropies, easy uniaxial anisotropy $K$ and hard plane anisotropy $K_\perp$,  
which originate from demagnetization. The aim is to analyze the impact of the longitudinal magnetic susceptibility under the 1D domain wall model, 
the demagnetization energy density could be written as
\begin{equation}
E_{\mathrm{an}} = -\frac{K}{M_e^2} M_x^2 + \frac{K_\perp}{M_e^2} M_z^2
\end{equation}
where $K=(1/2) (N_y-N_x)  \mu_0 M_e^2$ and $K_\perp=(1/2)  (N_z-N_y) \mu_0 M_e^2$.
In the limit case $\chi \rightarrow 0$ case, the effective anisotropy energy density $E_{\mathrm{an}}$ can be rewritten as 
\begin{equation}
E'_{\mathrm{an}} = K \sin^2 \theta (1+ \kappa \sin^2 \varphi),
\end{equation}
where $\mvec{m}=(\cos\theta,\sin\theta \cos\varphi,\sin\theta \sin\varphi)$ is used and $\kappa=K_\perp/K$ 
is the ratio between hard plane anisotropy strength and easy uniaxial anisotropy strength.

The dynamics of the domain wall with 1D profile can be described using 3 parameters \cite{OBrien2012}: the domain width $\Delta$, 
the domain wall position $q$ and  the domain wall tilt angle  $\phi$. In this domain wall model, one can assume that 
$\varphi(x,t)=\phi(t)$ is only a function of time. Thus, the magnetization profile for the head-to-head domain wall is given by
\begin{equation}
\theta(x,t)= 2 \tan^{-1} \exp\left(\frac{x-q(t)}{\Delta(t)}\right),\quad 
\varphi(x,t)=\phi(t).
\end{equation}

\begin{figure}[htbp]
\includegraphics[width=0.4\textwidth]{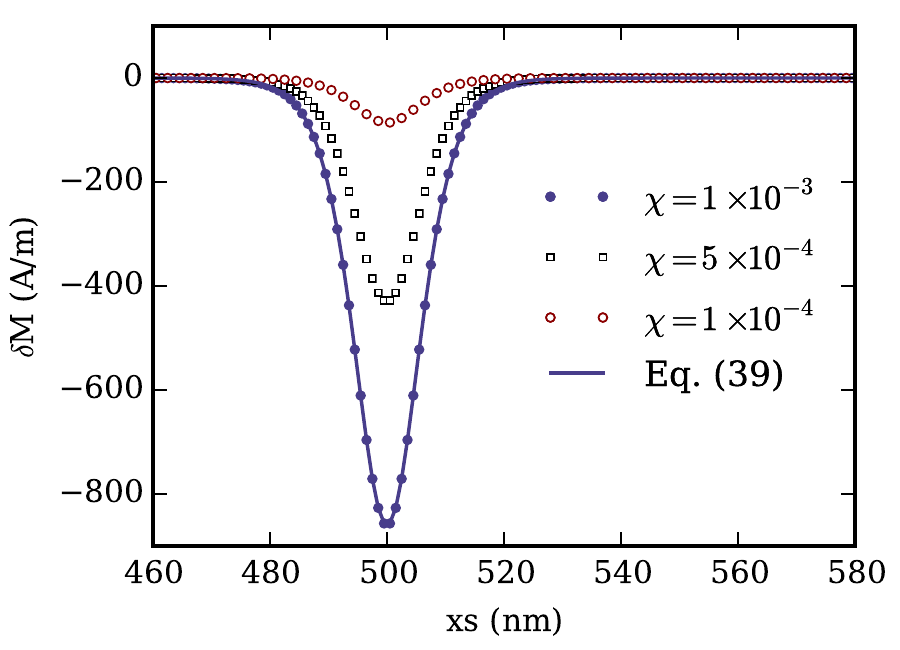} 
\caption{Simulation results of the magnetization length difference $\delta M$ for a 1D domain wall located at $x=500$ nm with 
$M_e = 8.6\times 10^5\,\mathrm{A/m}$ and $A=1.3\times 10^{-11}$ J/m.
The demagnetizing factors are selected to be $N_x=0$ and $N_y=N_z=0.5$.}
\label{fig_deltaM}
\end{figure}

Using the magnetization unit vector to calculate the exchange energy is a good approximation for the 
case $\chi \ll 1$, thus, the total energy density can be rewritten as 
\begin{equation}
E_{\mathrm{tot}}=\frac{\mu_0}{8\chi} \frac{(M^2-M_e^2)^2}{ M_e^2} + M^2 w_\mu(\mvec{m}),
\end{equation}
where 
\begin{equation}
w_\mu(\mvec{m})= \frac{A}{M_e^2} (\nabla \mvec{m})^2 -\frac{K}{M_e^2} m_x^2 + \frac{K_\perp}{M_e^2} m_z^2.
\end{equation}
Within the 1D domain wall profile, $H_\mathrm{m}$, the longitudinal component of the effective field is obtained:
\begin{equation}\label{eq_Hm}
H_\mathrm{m}=\mvec{m}\cdot\Heff= \frac{M}{2\chi M_e^2} (M_e^2 -M^2)- 2 M P \sin^2\theta
\end{equation}
where $P$ is defined as
\begin{equation}
P=\frac{1}{\mu_0M_e^2} \left[ \frac{A}{\Delta^2} + K (1+ \kappa \sin^2 \phi) \right].
\end{equation}
As we can see, $P$ is a function of the tilt angle $\phi$ and the domain wall width $\Delta$.
At the static state, 
$H_m$ should equal zero, i.e., ${d M}/{dt}=0$, which gives
\begin{equation}\label{eq.M}
M^2=(1-4\chi P \sin^2 \theta)M_e^2.
\end{equation}
Eq.~(\ref{eq.M}) shows that the difference between magnetization length $M$ and $M_e$
reaches its maximum at the center of the domain wall due to the effect of the exchange field, which also peaks in the centre of the domain wall.
According to Eq.~(\ref{eq.M}), we can estimate that the magnetization length difference is $\delta M \approx - 2 \chi P \sin^2 \theta$ for $\chi \ll 1$ case. 
Figure~\ref{fig_deltaM} shows the magnetization length differences of a 1D domain wall for various $\chi$, it can be seen that this 
approximation for $\delta M$ agrees very well with the simulation results. 

In the dynamic case, $H_m$ is not equal to zero.  If we wrote Eq.~(\ref{eq_Hm}) as $H_\mathrm{m} = F M$, we can find that
the nontrivial term that contributes to $H_m$ is
\begin{equation}
F =  \frac{1}{2\chi} (1-M^2/M_e^2)- 2  P \sin^2\theta.
\end{equation}
As an approximation for $H_\mathrm{m}$, we expect $d F/d t=0$~\cite{Sobolev1997_a}, 
which gives 
\begin{equation}\label{eq_Hm2}
H_{\mathrm{m}} =  \frac{4P}{\Delta} \frac{\chi}{\alpha} \frac{\dot{q}}{\gamma} m_t^2 m_x.
\end{equation}
In this approximation, we have ignored the terms containing $dP/dt$ and thus the amplitude of $H_{\mathrm{m}}$ is influenced by the domain wall velocity $\dot{q}$ only.
We employ the Lagrangian equation combined with dissipation function $\mathcal{F}$ to compute the 
domain wall dynamics \cite{BurkardHillebrands2006}. The Lagrange equations are 
\begin{equation}\label{eq.lag}
\frac{\partial \mathcal{L}}{\partial X} -\frac{d}{dt}\left( \frac{\partial \mathcal{L}}{\partial \dot{X}} \right)  + \frac{\partial \mathcal{F}}{\partial \dot{X}}=0, 
\end{equation}
where $X$ refers to $q$, $\phi$ and $\Delta$.
The dissipation function is defined by
$\mathcal{F}=\int \! F \, d x$
where
\begin{equation}\label{eq.F2}
F = \frac{1}{2}\mu_0 M_e \gamma [\alpha \Heff^2 + \sigma (\nabla \Heff)^2]
\end{equation}
is the dissipation density function. 

\subsection{Parallel relaxation}
We neglect the exchange damping term with assumption that $\sigma \ll \alpha \Delta^2$ and arrive at
\begin{equation}\label{eq.F}
F= \frac{1}{2}\alpha \mu_0 M_e \gamma  \Heff^2 = 
\frac{1}{2}\alpha \mu_0 M_e  \gamma (\mvec{H}_{\perp}^2+\mvec{H}_m^2).
\end{equation}
where $\mvec{H}_{\perp}$ and $\mvec{H}_m$ are the perpendicular and parallel components of the effective field.
If we also assume that $\alpha \sim \chi \ll 1$, $\mvec{H}_{\perp}^2$ can be approximated by Eq.~(\ref{eq_Heff}),
\begin{equation}\label{eq_h_perp}
\mvec{H}_{\perp}^2= \frac{1}{\gamma^2} \dot{\mvec{m}}^2 = \frac{1}{\gamma^2} (\dot{\theta}^2+\sin^2\theta \dot{\phi}^2).
\end{equation}
Substituting Eq.~(\ref{eq_Hm2}) and Eq.~(\ref{eq_h_perp}) into Eq.~(\ref{eq.F}) and integrating over space, we obtain
\begin{equation}\label{eq_Fchi}
\mathcal{F}=\frac{\alpha \mu_0 M_e}{\gamma} \left[ \dot{\phi}^2 \Delta +\frac{\dot{q}^2}{\Delta} (1+Q)\right],
\end{equation}
where we have ignored the $\dot{\Delta}$ term. This term leads to the optimal domain wall width~\cite{BurkardHillebrands2006}:
\begin{equation}\label{eq_delta}
\Delta = \sqrt{A/(K+K_\perp \sin^2 \phi)}
\end{equation}
and for $\kappa=0$ the optimal domain wall width reduces to $\Delta_0=\sqrt{A/K}$.
In what follows, the domain wall width parameter $\Delta(t)$ is approximated by the optimal wall width. 
The parameter $P$ is then given by
\begin{equation}
P=\frac{2K(1+ \kappa \sin^2 \phi)}{\mu_0M_e^2}=\frac{2}{\mu_0M_e^2}\frac{A}{\Delta^2},
\end{equation}
and it is straightforward  to find its minimum $P_0 = 2K/(\mu_0 M_e^2)$, which corresponds to $\Delta=\Delta_0$.

The introduced paramter $Q$ in Eq.~(\ref{eq_Fchi}) is given by $Q=(32/15)P^2(\chi/\alpha )^2$ 
and its value is determined by the ratio of $\chi$ and $\alpha$, which could be $\sim 1$ although we assume $\chi \sim \alpha \ll 1$.
Following the treatment of Ref. [\citenum{BurkardHillebrands2006}], the integrated Lagrangian action $\mathcal{L}$ is given by
\begin{gather}\label{eq.L}
\begin{split}
\mathcal{L}&=\int \! (E_\mathrm{tot} + \frac{\mu_0 M_e}{\gamma} \dot{\phi} \cos\theta)\, dx\\
&=\frac{2A}{\Delta}+2 \Delta K (1+ \kappa \sin^2 \phi)(1-V) \\
 &\qquad-2\mu_0 M_e H_a q +\frac{2\mu_0M_e}{\gamma}\dot{\phi} q,
\end{split}
\end{gather}
where $\mu_0 M_e \dot{\phi} \cos\theta/\gamma$ is the Berry phase term \cite{Shibata2011}, 
$V=8\chi P /3$ is a result of the varying magnetization that introduced a pinning potential. However, the potential is fairly small 
and therefore is negligible since $V \ll Q$. By substituting Eq.~(\ref{eq.L}) and Eq.~(\ref{eq.F}) into Eq.~(\ref{eq.lag}), 
\begin{gather}\label{eq_dqdp}
\begin{split}
&\dot{\phi}+\alpha \frac{\dot{q}}{\Delta}\left(1+Q\right ) = \gamma H_a,\\
&\frac{\dot{q}}{\Delta}-\alpha \dot{\phi}=\gamma \frac{H_k}{2} \sin2\phi.
\end{split}
\end{gather}
where $H_k=2 K_\perp/(\mu_0 M_e)$. The domain wall dynamics is governed by Eq.~(\ref{eq_dqdp}), 
 by eliminating $\dot{q}$ we obtain an equation about $\phi$,
\begin{equation}\label{eq_q}
\dot{\phi}=\frac{\gamma}{1+\alpha^2(1+Q)}\left [H_a-H_w (1 + Q) \sin2\phi \right]
\end{equation}
where $H_w=\alpha H_k/2$ is the Walker breakdown field. From Eq.~(\ref{eq_q}) we can find that the critical value of $\phi$ is approximately equal to $\pi/4$ 
if $Q \ll 1$, which leads to the maximum value of $P$ to be $P_1=2 K (1+\kappa/2)/(\mu_0 M_e^2)$.
There exists an equilibrium state $\phi^*$ such that $\dot{\phi}=0$ if $H_a<H_w(1 + Q)$,
\begin{equation}
\sin(2\phi^*)=h \equiv \frac{H_a}{H_w(1+Q)},
\end{equation}
which means the Walker breakdown field $H'_w$ for $\chi>0$  case is increased to $H_w (1 + \max \{Q\})$, i.e.,
\begin{equation}\label{eq.Hk}
H'_w=H_w \left[1+\frac{32}{15}P_1^2\left (\frac{\chi}{\alpha} \right )^2 \right],
\end{equation}
where $P_1$ is the maximum vlaue of $P$.
For this steady-state wall motion, the domain wall velocity is 
\begin{equation}\label{eq.v}
\dot{q}=\frac{\gamma H_a}{\alpha} \frac{\Delta^*}{1+Q(\Delta^*)},
\end{equation} 
where
\begin{equation}
\Delta^*=\Delta_0/\sqrt{1+\frac{\kappa}{2} (1-\sqrt{1-h^2})}.
\end{equation} 
Therefore, $\Delta^*  \rightarrow \Delta_0$ in the limit case $H_a \rightarrow 0$, and the domain wall mobility $\mu$ is given by
\begin{equation}
\mu= \frac{\gamma \Delta}{\alpha} \left[1+\frac{32}{15}P_0^2(\frac{\chi}{\alpha})^2 \right]^{-1} 
\end{equation} 
where $P_0$ is the minimum value of $P$. In Fig.~\ref{fig_dw} the corresponding Walker breakdown fields 
are plotted in vertical dash lines, which gives a good approximation for $\chi=5\times 10^{-4}$ and $\chi=1\times 10^{-4}$ cases.
The simulation results show that the Walker breakdown field $H_w$ could be changed significantly 
if the longitudinal susceptibility is comparable to the damping constant.

\subsection{Nonlocal damping}
\begin{figure}[tbhp]
\begin{center}
\includegraphics[width=0.4\textwidth]{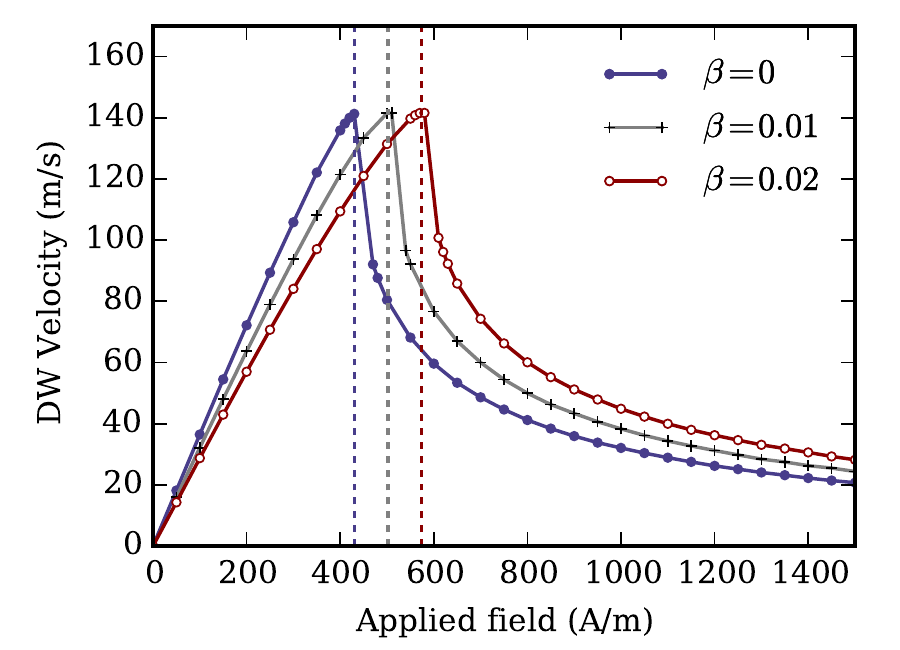} 
\caption{Simulations results of domain wall velocities for the limit case that $\chi\rightarrow 0$ with various exchange dampings. The parameters used are: 
$\alpha=0.005$, $N_y=0.4$ and $N_z=0.6$. The vertical dash lines are the breakdown fields computed 
with Eq.~(\ref{eq.Hk2}).}
\label{fig_dw_beta}
\end{center}
\end{figure}

In this part we consider the domain wall motion  influenced by exchange damping for the case 
that $\chi \rightarrow 0$. The dissipation density function
(\ref{eq.F2}) thus becomes
\begin{equation}\label{eq.F3}
F = \frac{1}{2}\mu_0 M_e \gamma \left[\alpha \mvec{H}_{\perp}^2 + \sigma (\nabla H_{\theta})^2+ \sigma (\nabla H_{\phi})^2 \right]
\end{equation}
where $H_{\theta}$ and $H_{\phi}$ are the two components of the effective field, and $\mvec{H}_{\perp}$ is 
computed using Eq.~(\ref{eq_h_perp}). After calculation we obtain
\begin{equation}
\mathcal{F}=\frac{ \mu_0 M_e}{\gamma} \left[ \dot{\phi}^2 (\alpha \Delta +\frac{1}{3} \frac{\sigma}{\Delta} ) + 
  \frac{\dot{q}^2}{\Delta} (\alpha+\frac{1}{3} \frac{\sigma}{\Delta^2})  \right].
\end{equation}
We take the same Lagrangian action (\ref{eq.L}) for $\chi=0$ and arrive at
\begin{gather}
\begin{split}
&\dot{\phi}+(\alpha+\frac{\sigma}{3 \Delta^2}) \frac{\dot{q}}{\Delta} = \gamma H_a,\\
\frac{\dot{q}}{\Delta}&-(\alpha+\frac{\sigma}{3 \Delta^2}) \dot{\phi}=\gamma \frac{H_k}{2} \sin2\phi.
\end{split}
\end{gather}
Similarly, the corresponding Walker breakdown field changes to 
\begin{equation}\label{eq.Hk2}
H'_w=\frac{1}{2} H_k \left(\alpha+ \frac{1}{3} \frac{\sigma}{\Delta^2_1} \right),
\end{equation}
where $\Delta_1=\Delta_0 \sqrt{1/(1+\kappa/2)}$. The domain wall mobility is given by
\begin{equation}\label{eq.mu2}
\frac{1}{\mu} = \frac{1}{\gamma \Delta_0} \left(\alpha+ \frac{1}{3} \frac{\sigma}{\Delta^2_0} \right).
\end{equation}
As we can see, the nonlocal damping term $\sigma$ influences the domain wall motion as well, and we 
can establish that $\sigma/\Delta^2=\beta (1+\kappa/2)K/(\mu_0 M_e^2) \propto \beta$.
Therefore, for the scenarios that $K\sim \mu_0 M_e^2$, the contributions from the Gilbert and nonlocal damping are of the order of
magnitude for both the domain wall mobility and Walker breakdown field.

Figure~\ref{fig_dw_beta} shows the domain wall velocities for domain wall motion driven by external fields in the limiting case of $\chi \rightarrow 0$.
The simulation results are based on a one-dimensional mesh with length $10000$ nm with cell size of $2$ nm. The damping $\alpha$ is 
set to $0.005$ and the demagnetizing factors are chosen to be $N_x=0$, $N_y=0.4$ and $N_z=0.6$. As predicted by Eq.~(\ref{eq.Hk2}),
the nonlocal damping $\beta$ leads to an increment of the Walker breakdown field, and Eq.~(\ref{eq.Hk2}) fits the simulation results very well.

\section{Summary}\label{section_summary}
We explain the ``exchange damping" in the Landau-Lifshitz-Baryakhtar (LLBar) equation as nonlocal damping 
by linking it to the spin current pumping, and therefore the LLBar (\ref{eq_bterm}) can be considered 
as a phenomenological equation to describe the nonlocal damping.
In the presence of nonlocal damping, the lifetime and propagation length of short-wavelength magnons could be much
shorter than those given by the LLG equation. Our simulation results show that the spin wave amplitude decays much faster in the presence of nonlocal damping 
when spin waves propagate along a single rod. The analytical result shows that there is extra nonlinear dependence scaling with $k^2$ 
between $\lambda k$ (the product of spin wave decay constant $\lambda$ and wave vector $k$) and 
frequency $\omega$ due to the nonlocal damping.
Using the micromagnetic simulation based on the LLBar equation, we show that the difference between magnetization 
length $M$ and $M_e$ reaches its maximum at the center of the domain wall. For the cases that $\chi \sim \alpha$ where
$\chi$ is the longitudinal magnetic susceptibility and $\alpha$ is the Gilbert damping, the Walker breakdown field
will increase significantly.
By using a 1D domain wall model, we also show that both the domain wall mobility and the Walker breakdown field
are strongly influenced by the nonlocal damping as well.

\acknowledgments
We acknowledge the financial support from EPSRC's DTC grant EP/G03690X/1. 
W.W. thanks the China Scholarship Council for financial assistance.
The research leading to these results has received funding from the European Community's Seventh 
Framework Programme (FP7/2007-2013) under Grant Agreement n247556 (NoWaPhen) and from the European 
Union's Horizon 2020 research and innovation programme under the Marie Sk\l{}odowska-Curie grant agreement No 644348 (MagIC).

\appendix

\section{Derivation of equation (\ref{eq_llb})}\label{app_llb}
We split the perpendicular spin current $\mvec{j}_{\perp,i}$ into two components, 
\begin{equation}
\mvec{j}_{\perp,i} = \mvec{j}_i^a + \mvec{j}_i^b,
\end{equation}
where we write $\lambda_e/\gamma$ as $\tilde{\sigma}$,
\begin{eqnarray}
\mvec{j}_i^a = -\tilde{\sigma} ( \partial_i \mvec{m} \times  \partial_t \mvec{m})\\
\mvec{j}_i^b = -\tilde{\sigma} ( \mvec{m} \times  \partial_i\partial_t \mvec{m}) 
\end{eqnarray}
The  torque $\bm{\tau}_a$ generated by spin current $\mvec{j}_i^a$ is given
by $\bm{\tau}_a =(\partial_i \mvec{j}_i^a )_{\perp}$, i.e.,
\begin{equation}\label{eq_torque_a}
\bm{\tau}_a= \tilde{\sigma} \mvec{m} \times [ \partial_i\mvec{m} \times  (\partial_t \mvec{m} \times  \partial_i \mvec{m})] 
\end{equation}
where we have used the identities
$\mvec{m}\cdot \partial_i \partial_t \mvec{m}=-\partial_i \mvec{m} \cdot \partial_t \mvec{m}$ and 
$\mvec{m}\cdot \partial_i \partial_i \mvec{m}=-\partial_i \mvec{m} \cdot \partial_i \mvec{m}$. 
Meanwhile, the corresponding torque $\bm{\tau}_b$ can be computed by 
$\bm{\tau}_b =(\partial_i \mvec{j}_i^b )_{\perp}$, which gives
\begin{equation}
\bm{\tau}_b = \bm{\tau}_a - \tilde{\sigma} (\partial_i \mvec{m} \cdot \partial_i \mvec{m}) \mvec{m} \times \partial_t \mvec{m}
 - \tilde{\sigma} \mvec{m} \times \nabla^2  \partial_t \mvec{m}
\end{equation}
Note that $\bm{\tau}_a=\tilde{\sigma} \partial_i \mvec{m} [(\partial_t \mvec{m} \times  \partial_i \mvec{m})\cdot \mvec{m}]$
can be changed into the tensor form, 
\begin{equation}
\bm{\tau}_a = \mvec{m}\times (\mathcal{D}^0 \cdot \partial_t \mvec{m}), 
\end{equation}
where 
\begin{equation}\label{eq_D0}
\mathcal{D}^0_{\alpha\beta} = \tilde{\sigma} (\mvec{m}\times \partial_i \mvec{m})_\alpha (\mvec{m}\times \partial_i \mvec{m})_\beta.
\end{equation}
Therefore, we obtain for $\bm{\tau}_a +\bm{\tau}_b$,
 \begin{equation}
 \bm{\tau}_a +\bm{\tau}_b = 
 \mvec{m}\times (\mathcal{D} \cdot \partial_t \mvec{m}) 
 - \tilde{\sigma} \mvec{m} \times \nabla^2  \partial_t \mvec{m}
\end{equation} 
where $\mathcal{D}$ is a $3\times 3$ tensor,
\begin{equation}\label{eq_Dtensor2}
\mathcal{D}_{\alpha\beta} = 2 \tilde{\sigma} (\mvec{m}\times \partial_i \mvec{m})_\alpha (\mvec{m}\times \partial_i \mvec{m})_\beta
- \tilde{\sigma} (\partial_i \mvec{m} \cdot \partial_i \mvec{m}) \delta_{\alpha \beta}.
\end{equation}

\section{Derivation of equation (\ref{eq_sw})} \label{app_sw}
We introduce a new variable $\mvec{s}$ to represent the second term in the (\ref{eq_sw_d}), 
i.e., $\mvec{s} = \mvec{m}_0 e^{i( \tilde{k} x - \omega t)}$, so we have
\begin{eqnarray}
&\mvec{m}= \mvec{e}_\mathrm{x}+\mvec{s},\\
&\frac{d \mvec{m}}{d t}=-i \,\omega \,\mvec{s},\\
&\Heff = H_s (1+s'_x) \mvec{e}_\mathrm{x}- D \tilde{k}^2 \mvec{s}
\end{eqnarray}
where $s'_x \approx (1/2) (s_x^2-s^2)$. 
Considering the fact $|\mvec{s}| \ll 1$ and neglect the high order term $s^2$, one 
obtains $\Heff^\perp= -  (H_s + D \tilde{k}^2 ) \mvec{s} $ and thus
\begin{equation}
\Heff^b = c\, \mvec{e}_\mathrm{x}+ d \,\mvec{s}
\end{equation}
where
\begin{eqnarray}
&c= \alpha H_s (1+ s'_x),\\
&d=-\beta G  \tilde{k}^2(D \tilde{k}^2 +H_s) - \alpha D \tilde{k}^2
\end{eqnarray}
Substituting the above equations into (\ref{eq_bterm}), we have
\begin{equation}
\frac{i \omega}{\gamma}  \left[ \begin{array}{c}
s_x\\s_y\\s_z
\end{array} \right]
= f \left[ \begin{array}{c}
  0\\
   s_z \\
  - s_y
\end{array} \right]
+(c-d)
 \left[ \begin{array}{c}
-(s_y^2 + s_z^2) \\
 (1 + s_x) s_y \\
(1 + s_x) s_z
 \end{array} \right],
\end{equation}
where $f =  H_s (1+s'_x) + D \tilde{k}^2$.
Neglecting high order terms such as $s_x^2$ and $s_x s_y$ we obtained,
\begin{equation}\label{eq_appendix_matrix}
 \begin{bmatrix}
  \gamma(\alpha H_s-d) -i \omega &  \tilde{w}_0 \\
  - \tilde{w}_0 &  \gamma(\alpha H_s-d) -i \omega 
 \end{bmatrix}
  \begin{bmatrix}
 s_{y} \\
  s_{z} 
 \end{bmatrix} =
\begin{bmatrix}
0 \\
 0 
 \end{bmatrix}.
\end{equation}
Therefore, Eq.~(\ref{eq_sw}) can be 
obtained by setting the determinant of the matrix in (\ref{eq_appendix_matrix}) to zero.

\bibliographystyle{apsrev4-1}
%

\end{document}